%% file: main.tex
\documentclass[lettersize, journal]{IEEEtran}

\usepackage[utf8]{inputenc}

\usepackage{flushend}
\usepackage{multirow}
\usepackage{xcolor}
\usepackage{balance}
\usepackage{caption}
\usepackage{subcaption}
\usepackage{graphicx}
\usepackage{comment}
\usepackage{amsfonts}
\usepackage{amsthm}
\usepackage{xcolor}
\usepackage{floatrow}
\DeclareFloatFont{tiny}{\tiny}
\begin{document}

\title{Neural Fuzzy Extractors: A Secure Way to Use Artificial Neural Networks for Biometric User Authentication}

 \author{Abhishek Jana, 
Bipin Paudel,
Md Kamruzzaman Sarker,
Monireh Ebrahimi,
Pascal Hitzler and \\ George T Amariucai

\thanks{Abhishek Jana is with the Department of Pyisics at Kansas State University, e-mail: ajana@ksu.edu.}
\thanks{Bipin Paudel, Md Kamruzzaman Sarker, Monireh Ebrahimi, Pascal Hitzler and George T Amariucai are with the Department of Computer Science at Kansas State University. e-mails: bipinp@ksu.edu, mdkamruzzamansarker@ksu.edu, monireh@ksu.edu, hitzler@ksu.edu, amariucai@ksu.edu.}
}

\maketitle


\begin{abstract}
Powered by new advances in sensor development and artificial intelligence, the decreasing cost of computation, and the pervasiveness of handheld computation devices, biometric user authentication (and identification) is rapidly becoming ubiquitous. Modern approaches to biometric authentication, based on sophisticated machine learning techniques, cannot avoid storing either trained-classifier details or explicit user biometric data, thus exposing users’ credentials to falsification. In this paper, we introduce a secure way to handle user-specific information involved with the use of artificial neural networks for biometric authentication. Our proposed architecture, called a Neural Fuzzy Extractor (NFE), allows the coupling of pre-existing classifiers with fuzzy extractors, through an artificial-neural-network-based buffer called an expander, with minimal or no performance degradation. The NFE thus offers all the performance advantages of modern deep-learning-based classifiers and all the security of standard fuzzy extractors. We demonstrate the NFE retrofit of a few classic artificial neural networks, for simple biometric authentication scenarios.
\end{abstract}



   
\maketitle  

\input{intro.tex}

\input{related-works.tex}

\input{extractor.tex}

\input{classifier.tex}

\input{evaluation.tex}

\input{conclusion.tex}

\input{acknowledgement.tex}
\bibliographystyle{IEEEtran}
\bibliography{refs,george,abhishek}

\end{document}

%% file: intro.tex
\section{Introduction} \label{intro}

Secure architectures for password-based authentication avoid storing the passwords corresponding to each username, and resort instead to storing cryptographic hash digests of such passwords (along with salt, pepper, and other such auxiliary randomness). The rationale behind this widely-accepted paradigm is that not even a superuser, with complete access to the entire file system, should be able to fraudulently log in as one of the other, less privileged users of the system.

When using standard biometric authentication in place of (or in addition to) a password, similar functionality can be achieved through the use of fuzzy extractors \cite{dodis2004fuzzy}. However, despite their security guarantees, fuzzy extractor architectures are rarely deployed in practice. 
In particular, in the case of fingerprint-based authentication it turns out that similarity-score algorithms usually perform better than fuzzy extractors, and this has led to a preference for the former. For example, using fuzzy extractors, \cite{arakala2007fuzzy} reports an equal error rate of 15\% on the FVC2000 database \cite{fvc2000}, while the top five FVC2000 competitors report equal error rates between 5\% and 0.7\% \cite{maio2002fvc2000}.

However, using similarity-score algorithms involves the storing of fingerprint databases, which are vulnerable to leakage. Notable examples are the Office of Personnel Management data breach  of 2015 \cite{gootman2016opm}, in which 5.6 million sets of fingerprints were leaked, and the more recent Suprema Biostar leak of 2019 \cite{suprema2019}, which compromised the fingerprint and facial biometric information of more than one million people. An authentication method which has both the security of fuzzy extractors and the superior performance of similarity-score algorithms would clearly be prefered. 

The reason behind the inferior performance of fuzzy extractors may be traced back to the very essence of fuzzy extractor functionality, the first stage of which consists of a channel decoding mechanism over a vector space. By nature, good channel coding implicitly assumes spherical (or close to spherical) decoding regions, which correspond to the decision regions of the vector-space-based classifier. In contrast to this, normal support-vector machines (SVMs) can learn highly irregular decision regions -- hence their superiority. Similarly, artificial neural networks (ANNs) are known to outperform even the best of SVMs (at least in situations in which training data is abundant), and this is attributed to their ability to learn highly-irregular classification functions.

Unfortunately, both SVMs and ANNs (as well as the other frequently-used classifiers, like k-nearest neighbors (KNN), decision trees and random forests, etc.) rely on learned structures that have to be stored in non-volatile memory, similarly to a password file. A malicious user, with access to this information, could use the learned structure (for example, by back-tracking through an ANN, or by simply choosing a vector in the proper decision region, for an SVM) to produce synthetic inputs guaranteed to pass the authentication test.

The question that arises naturally is then how we can protect a user's biometric authentication information in a manner similar to the way in which we treat passwords, but without suffering from the spherical restrictions of the fuzzy extractors. In this paper, we propose a new (and severely overdue) such architecture, which we call \emph{neural fuzzy extractors (NFEs)}.

\begin{figure}[b]
	\centering
	\includegraphics[width=\linewidth]{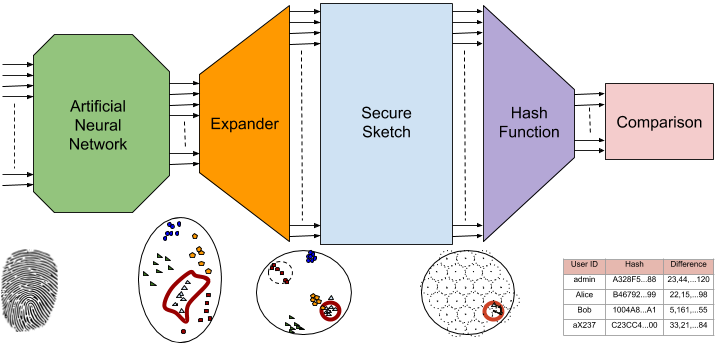}
	\caption{Neural Fuzzy Extractor architecture.} \label{architecture}
	\vspace{-4mm}
\end{figure}

NFEs are a concatenation of a classifier -- such as an artificial neural network -- with a fuzzy extractor, as illustrated in Figure \ref{architecture}. Nevertheless, as most ANNs are designed to output class labels or scores contained within some subspace of a certain vector space, and we want our NFE construction to be as close as possible to an add-on, to take advantage of many already-existing and well performing classification architectures, we have to retrofit NFEs to regular ANNs. We do this by constructing an interface between the ANN and the fuzzy extractor's decoding mechanism, which we call an \emph{expander}. An expander will typically consist of an additional ANN with a few layers, that can be added to the end of any type of ANN, and the sole purpose of which is to re-cast the output embedding of the original ANN to a vector space in which representatives from each class cluster together in sphere-like clusters. The expander may be trained independently, based on labeled embedings from the original ANN, or together with the original ANN.

Of course, some already-existing ANN architectures may be naturally suited to concatenation with fuzzy extractors, in the sense that their output embeddings corresponding to different classes are already in a vector space and already cluster in spheres. In these cases, no additional expander is necessary.

We should also note that the above-mentioned vector spaces need not be defined over $\mathbb{R}^n$ -- as they are for the particular architecture discussed in this paper -- but can also be defined over any other field, like $GF(2^n)$. The latter choice is easier to deal with from the perspective of the secure sketch, mainly due to the wide availability of binary channel capacity-achieving error correction codes (ECCs) -- by contrast, additive white Gaussian noise (AWGN) channel capacity-achieving ECCs are few and suffering from very complex decoding mechanisms. On the other hand, forcing the output of the expander into binary vectors may incur unacceptable performance losses, which is why we leave the investigation of binary-ECC-based NFEs to future work.

The contributions of this paper are as follows.
\begin{enumerate}
    \item We introduce NFEs, a first secure architecture for handling ANN-based biometric user authentication.
    \item We show that NFEs can be retro-fitted to work with most of the already-existing ANN-based biometric authentication architectures.
    \item We demonstrate our construction on three already-existing fingerprint-based authentication architectures, and we show that the NFE retrofit has marginal, if any, effects on performance, while introducing non-negligible but acceptable overhead to the training and running times.
\end{enumerate}

The remainder of the paper is organized as follows. Section~\ref{relatedwork} provides a brief survey of already-existing biometric authentication protocols, focusing mostly on recent results involving artificial neural networks, and techniques that use fuzzy extractors. Section \ref{fuzzy} introduces the NFE architecture and provides some general implementation insights. Section \ref{demo} presents the design decisions involved in our instantiation of the NFE architecture, namely the types of classifiers considered, the architecture of the expander and the choice of the error-correction code. Section \ref{eval} describes the decisions involved in the design of our evaluation framework, such as the datasets and the experimental setup, and presents our results in terms of authentication performance, training and running times. Finally, conclusions are drawn in Section \ref{conclusions}. 

%% file: related-works.tex
\section{Related Work}\label{relatedwork}
\subsection{Biometric Authentication and Identification}
The literature contains many early attempts at leveraging the ANN capabilities for biometric verification and authentication. \cite{brady1999biometric} has filed a patent on biometric recognition using a classification neural network. The patented biometric recognition system involves two phases: creation of a master pattern set of authorized users' biometric identifications and authentications using a classification neural network. \cite{allah2005artificial} developed a new supervised recurrent neural network for fingerprint authentication. Their approaches used similarity measures of features for clustering and ranking of the fingerprint representations stored in their database. \cite{wong2001enhanced} used both artificial neural networks and k-nearest neighbors as possible classifiers for typing pattern identification. 
\cite{yong2004weightless} investigates the implementation of Weightless Neural Networks (WNNs) as a pattern recognition tool to classify users’ typing patterns and thus attempts to separate the real users from impostors. \cite{nazeer2007face} used artificial neural networks for face representation learning and recognition.

 With the recent resurgence of interest in Deep Learning models, in recent years we have witnessed significant progress in representation learning for biometric identifiers by deep neural networks. \cite{tang2017fingernet} has proposed FingerNet, a unified deep network for fingerprint minutiae extraction.
They propose a new way to design a deep convolutional network combining domain knowledge and the representation ability of deep learning. In terms of orientation estimation, segmentation, enhancement and minutiae extraction, several typical traditional methods that performed well on rolled/slap fingerprints are transformed into a convolutional approach and integrated as a unified plain network. \cite{darlow2017fingerprint} posed minutiae extraction as a machine learning problem and proposed a deep neural network -- MENet, for Minutiae Extraction Network -- to learn a data-driven representation of minutiae points. \cite{menotti2015deep} used deep representations for Iris, Face, and Fingerprint Spoofing Detection. Similarly, \cite{engelsma2019fingerprints} learned fingerprint representations. \cite{jeon2017fingerprint} proposed three variations of the VGGNet structure for fingerprint classification. 

Similarly, \cite{hammad2018multimodal} proposed a secure multimodal biometric system that uses a convolutional neural network (CNN) and a Q-Gaussian multi support vector machine (QG-MSVM) based on different level fusion. They developed two authentication systems with two different level fusion algorithms: a feature level fusion and a decision level fusion. The feature extraction for individual modalities is performed using a CNN. In this step, they selected two layers from the CNN that achieved the highest accuracy, in which each layer is regarded as a separate feature descriptor. After that, they combined them using the proposed internal fusion to generate the biometric templates. In the next step, they applied one of the cancelable biometric techniques to protect these templates and increase the security of the proposed system. 

Likewise, \cite{nayak2012multimodal} conducted multimodal biometric face and fingerprint recognition using neural networks based on adaptive principal component analysis and multilayer perceptrons. \cite{stojanovic2017novel}  proposed a novel latent overlapped fingerprints separation algorithm based on neural networks. \cite{nogueira2016fingerprint} used convolutional neural networks (CNNs) for fingerprint liveness detection.

\cite{page2015utilizing} leveraged neural networks to both identify QRS complex segments of ECG signals and then performed user authentication on these segments. \cite{mai2011ecg} used multilayer perceptrons and radial basis function neural networks for electrocardiogram (ECG) biometric authentication. \cite{salloum2017ecg} proposed the use of various recurrent neural network (RNN) architectures (including vanilla, long short-term memory (LSTM), gated recurrent unit (GRU), unidirectional, and bidirectional networks) for ECG-based biometrics identification/classification and authentication. \cite{labati2019deep} presents Deep-ECG, a CNN-based biometric approach for ECG signals identification, verification and periodic re-authentication. Deep-ECG extracts prominent features from one or more leads using a deep CNN and compares biometric templates by computing simple and fast distance functions for verification or identification. \cite{el2016face} showed the novel use and effectiveness of deep learning CNN architectures for automatic rather than hand-crafted feature extraction for robust face recognition across time lapses. They show CNNs using the VGG-Face deep networks produce highly discriminative and interoperable features that are robust to aging variations even across a mix of biometric datasets.

\subsection{Fuzzy Extractors}

Fuzzy extractors were introduced in \cite{dodis2004fuzzy} as a secure way of coping with user biometrics -- for which every new entry is slightly different from previous ones, but all entries share some common main features. The idea was that, instead of storing representative entries,  for direct comparison to the new entries upon authentication request, the system should only store digests obtained through cryptographic hash functions -- thus preventing biometric falsification.

The idea was quickly adapted to various types of biometric authentication mechanisms, like those based on fingerprints \cite{arakala2007fuzzy,yang2012delaunay,xi2011alignment,li2008fuzzy,tong2007biometric}, iris scans \cite{marino2012crypto,hernandez2009biometric,bringer2007optimal}, face \cite{sutcu2009design} or gait \cite{hoang2014secure,hoang2015gait}.

More recently, fuzzy extractors were used in the context of more sophisticated and specialized secure authentication mechanisms, like the one in \cite{das2017secure}, designed specifically for wireless sensor networks (like body-area networks), or the ones in \cite{delvaux2016efficient, liao2017impact}, which deal with the outputs of physically-unclonable functions (PUFs).

However, to the best of our knowledge, at the time of this writing, no works exist on the application of fuzzy extractors on biometric data pre-processed by sophisticated classifiers like artificial neural networks.

\subsection{Privacy-Preserving Biometrics}

Our problem is related to the problem of privacy-preserving biometric authentication \cite{tran2021biometrics,belguechi2011overview}, to the extent that our NFE solution helps with some of the common objectives of privacy-preserving biometrics, such as biometric template protection \cite{belguechi2011overview}. Nevertheless, the problem of privacy-preserving biometric authentication is much broader, dealing with the privacy of the entire biometric authentication process, from the security of the channel between sensor and authentication server, to the security of the authentication database, and from anonymous biometric verification \cite{evans2011efficient} to biometric cancellability \cite{teoh2008cancellable}. The techniques employed in the literature to achieve these diverse goals range from the use of fuzzy extractors \cite{das2018biometrics} to secure multi-party computation \cite{bringer2013privacy}, and from zero-knowledge proofs of knowledge \cite{bhargav2007privacy} to fully-homomorphic encryption \cite{kim2020efficient,evans2011efficient}.

%% file: extractor.tex
\section{Preliminary Concepts: the Architecture of a Neural Fuzzy Extractor}\label{fuzzy}

Most classifiers -- whether neural networks or vector-space-based -- will output a vector representation of the input data. In some cases, the output vectors are already appropriate for direct input to the secure sketch (see Figure \ref{architecture}). To satisfy this property, the classifier outputs corresponding to each class of interest have to cluster in a somewhat spherical region of the vector space. This is because the secure sketch will use codes designed for error correction on either white Gaussian noise (AWGN) channels, or on binary symmetric channels (BSCs), where the decoding region is spherical by construction. So if the classifier's decision regions are not already spherical, imposing spherical decoding regions on top of them will invariably degrade the classification performance.

Unfortunately, in most cases, the classifiers are solely designed for good accuracy, without extra constraints on the spherical shape of their decision regions. In such cases, we propose the use of an \emph{expander}, as illustrated in Figure \ref{architecture}, to further shape the classifier's decision regions into spherical ones. We call this procedure \emph{retrofitting the NFE to pre-existing classifiers}. Intuitively, we expect that in order to avoid reducing the overall accuracy, we need to preserve as much of the information content of the classifier output as possible. Intuitively, this means that the expander should generally project the vector representations of the classifier's output to a larger-dimensional space (hence the term ``expander''), in which the decision regions can be made spherical without significant accuracy penalties. We should note here that in cases in which the classifier is a neural network, the last layer often reduces the dimension of the data -- for instance, to fit the number of relevant classes. In such cases, we propose to remove the last (low-dimensional) layer of the neural network before attaching the expander. In the cases in which the neural network presents an extremely wide output, the expander will in fact need to act as a compression mechanism, in order to reduce the dimension of the network's output layer to one that is tractable by the secure sketch -- as we shall see shortly, the decoding procedure employed by the secure sketch can be quite time consuming \cite{liu2005decoding} (albeit running in time linear in the code length).

\subsection{The Classifier}\label{classifier}

As mentioned above, with the option of retrofitting the NFE to pre-existing classifiers, the only requirement for the classifier is to process the users' biometric readings into vectors, in a vector space in which classification is possible with reasonable accuracy. We do not expect that retrofitting with the NFE will improve this accuracy in any way -- nor is that the purpose of the retrofit. The best we can hope for is that the original classifier's accuracy is maintained, while the security of the system is greatly improved. As such, our NFE scheme can work with multiple types of already-existing classifiers, for example K-nearest-neighbors, support-vector machines, or neural networks. The ``\emph{neural}'' part of ``NFE'' refers not to the retrofitted classifier, but rather to the expander, which is invariably implemented as an artificial neural network.

\subsection{The Expander}\label{expander}

The expander will be constructed as a neural network, will take as input the output of the (trimmed or intact) original classifier, and will be trained using a cost function that penalizes deviations form a spherical shape. If the original classifier is also a neural network, the training of the expander can be done at the same time as that of the original classifier -- in essence, the procedure consists of simply adding a constraint on the shape of the decision regions. However, when retrofitting the NFE to an already-trained neural network, separate training of the expander can be accomplished by feeding it with labeled outputs of the (trimmed or intact) original network, corresponding to some (original or novel) training dataset.   

\subsection{The Secure Sketch}\label{sketch}

The secure sketch, as defined in \cite{dodis2004fuzzy}, consists of (1) a mechanism for mapping the fuzzy biometric of a user to a fixed point in the vector space of the biometric representation, coupled with (2) a secure method for storing such identifying user information. If the fuzzy biometric data of our user is situated in a roughly spherical region of the vector space, then the first part can be accomplished by defining an error-correction code over the vector space, and shifting it so that the user's decision sphere overlaps with one of the decoding regions, corresponding to one of the codewords.

\begin{figure*}[]
	\centering
	\includegraphics[width=\linewidth]{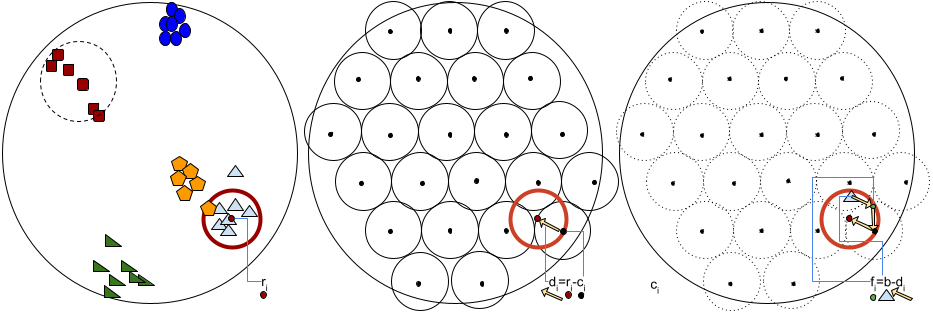}
	\caption{Secure sketch construction. Left: The large circle is chosen to contain all available embeddings, from all users; the small circle for the authentic user (AU) is chosen to yield a favorable false positive-false negative compromise (different radii may be chosen for different AUs). Middle: A codebook is constructed, with Voronoi region congruent to AU's decision region; the center of AU's decision region ($r_i$) is decoded to the closest codeword ($c_i$), and the difference between the center of AU's decision region and this codeword ($d_i$) is saved to the AU's record along with the hash of the center of AU's decision region. Right: The AU submits a new sample for authentication ($b$); by subtracting the difference vector ($d_i$) and decoding to the nearest codeword ($c_i$), the previously-identified codeword is recovered. We then add the difference to the recovered codeword (again $d_i$), and obtain the center of AU's decision region ($r_i$); its hash is compared to AU's record.} \label{codebook}
	\vspace{-4mm}
\end{figure*}

This construction is described intuitively in Figure \ref{codebook}. In the left-most part of the figure, we can see that the authentic user's (AU's) biometric data-points (represented as light blue triangles) register (mostly) inside the bottom-right red sphere -- which is the user's \emph{decision region}. The radius of this sphere is user-specific, and has to be chosen to provide a good compromise between false positives and false negatives. The large circle is chosen to contain all available embeddings, from all users; in the general multi-dimensional case, this circle corresponds to the hyper-sphere that represents the support of the expander's output -- basically, the region of the vector space in which one would expect to find data points corresponding to the embeddings of any user's biometric data.

\paragraph*{Registration phase} For the registration phase, a codebook is defined over the vector space, and restricted to the support of the expander's output. Different methods can be used to define such a code, but an optimal choice would be a capacity-achieving code (for an additive White Gaussian noise (AWGN) channel, if the vector space is of the form $\mathbb{R}^n$, and for a binary symmetric channel if the vector space is defined over $GF(2^n)$). For our specific example, in which embeddings are defined over  $\mathbb{R}^{128}$, we choose a low-density lattice code (LDLC) \cite{sommer2008low} in $128$ dimensions. This codebook is made up of a set of codewords, which are represented by the black dots in the middle portion of Figure \ref{codebook}. Each such black dot is surrounded by its decoding sphere (or \emph{Voronoi} region). Any data point (represented as a vector) falling inside a particular codeword's Voronoi region can be decoded to this particular codeword (which is that codeword in the codebook that is closest to the data point).

Next, we identify the center of the AU's decision region -- denote it as $\mathbf{r}_i$ -- and ``decode'' it to the closest codeword in the codebook -- denote this codeword as $\mathbf{c}_i$. We then calculate the difference vector (DV) $\mathbf{d}_i=\mathbf{r}_i-\mathbf{c}_i$ between the center of the AU's decision region and the closest codeword (a translated copy of $\mathbf{d}_i$ is shown as the small yellow arrow in the bottom-right part of the  middle part of Figure \ref{codebook}). The DV $\mathbf{d}_i$ is stored as part of the AU's authentication record, along with the hash $h(\mathbf{r}_i)$ of the center of the AU's decision region.

\paragraph*{Verification phase} For each of registered user $i$, the verifier has access to their tuple $(\mathbf{d}_i, h(\mathbf{r}_i))$. Upon verifying a user's claim to be user $i$, the current biometric reading -- denote it as $\mathbf{b}$ is placed (by virtue of the expander-enhanced classifier) into the vector space, and the DV is subtracted from it. Let the result be $\mathbf{f}_i=\mathbf{b}-\mathbf{d}_i$. In the right part of Figure \ref{codebook}, $\mathbf{b}$ is represented as a light-blue triangle, while $\mathbf{f}_i$ is the tip of the upper yellow arrow. The vector $\mathbf{f}_i$ is then decoded to the closest codeword -- denote this codeword as $\mathbf{c}_j$ and note that if $\mathbf{b}$ is indeed the biometric of user $i$, then we should have $\mathbf{c}_j=\mathbf{c}_i$. Finally, the DV $\mathbf{d}_i$ is added to this codeword $\mathbf{c}_j$ (in an attempt to recover the center of the AU's decision region), and the hash of the result $h(\mathbf{c}_j+\mathbf{d}_i)$ is compared to the one in user $i$'s record $h(\mathbf{r}_i)$. The user is authenticated if $h(\mathbf{c}_j+\mathbf{d}_i)=h(\mathbf{r}_i)$.

\emph{NOTE 1}: In general, instead of \emph{decoding} the center of the AU's decision region to the closest codeword, we could simply choose a random codeword, and calculate the DV between the center of the AU's decision region and this codeword. But in this case, additional steps have to be taken to ensure that the DV does not leak any information about the center of the AU's decision region. For example, if we choose a codeword in the upper-left of the large sphere in  Figure \ref{codebook}, the DV has a large amplitude, and an attacker could infer that the center of the AU's decision region is in the lower right. To avoid such leakage, we would need to add to DV an additional random vector, the effect of which is neutral when wrapped around the large sphere (zero modulo the large sphere).

\emph{NOTE 2}: It may appear at a first glance that the system can be further simplified by hashing directly the (closest, or randomly-chosen) codeword. However, with such an implementation, care must be taken to ensure that different users are assigned different codewords, which results in a net increase of the system complexity. Using the extra steps required to store the hash of the center of the AU's decision region as above will naturally ensure that different users have different hashes in the authentication table.

\subsection{The Hash}\label{hash}

The hash component of the NFE architecture is implemented as a simple cryptographic hash function, to be chosen according to the most recent NIST recommendations. At the time of this writing, hash functions from the SHA-2 and \mbox{SHA-3} families would be perfectly adequate. The user authentication database stores, along with user names, the following user identifying components: (1) codebook parameters (or just a decoding algorithm), (2) difference vector (DV), as explained in Section \ref{sketch} above; (3) any non-secret randomness such as salt and pepper used during the hashing process, and (4) the hash value (with salt, pepper, etc.) of the center of AU's decision region, as explained in Section \ref{sketch} above.

\subsection{Security Considerations}\label{security}

\paragraph*{Threat Model}
We assume a biometric-based authentication system, implemented with the help of an artificial neural network on an authentication server (which, as a particular case, and in a broad sense, can be hosted on the local machine to which users attempt to log in). Each of the legitimate users of the system undergo the registration phase, in which multiple readings of their biometric feature(s) are taken and used to train the authentication system. The trained authentication system, as well as the users' authentication records, are stored (possibly encrypted) on the authentication server. We assume no malicious interference during, or eavesdropping of, the registration phase -- this phase can be completed in a protected environment, just like the one required when users set up their new accounts and passwords on any password-based authentication system.
Further, we assume that the attacker cannot observe an authentication procedure in real time, and hence does not have direct access to the biometric information entered by the user. In reality, such access is possible if the attacker eavesdrops the connection between authentication server and biometric sensor, or if the attacker has admin privileges and scans the memory during the authentication process. We should note that all of these assumptions are standard in biometric authentication, and any efforts to strengthen the authentication system for the event when any of these assumptions fail are compatible with, but outside the scope of, our work.

The threat model consists of an attacker who can recover the weights of the ANN-based authentication system, as well as all users' biometric records. The attacker has access to this information only after the registration phase is complete, and has no further access to the authentication server, except potentially as a regular legitimate user of the system. Further, the attacker has ready access to technology that can produce exactly one biometric reading that is fully under the control of the attacker, and cannot be distinguished from a real biometric reading from a real person. The restriction to exactly one biometric reading is required to avoid the situation when the attacker brute-forces the biometric authentication system -- the effort required by a biometric brute-force attack depends only on the entropy of the biometric, and has little to do with the authentication mechanism that is using it, and the security of which this paper aims to improve.

The attacker's goal is to complete a successful login attempt as any one of the system's legitimate users -- of course, with the exception of the user that the attacker already owns or has compromised. The attacker wins if they can log in as a (different) legitimate user of the system, with probability significantly higher than a non-legitimate random person attempting to log in by properly providing biometric readings to the authentication system.

\paragraph*{Security Evaluation of the NFE}
In order to log in as a (different) legitimate user of the system, the attacker would have to input the user name of a user of their choice (the attacker has a list of all user names from the authentication records), and to produce a biometric reading which, when passed through the authentication mechanism, yields the same output as the one in the authentication record associated to the chosen user name.

In the case of a standard ANN-based authentication mechanism, this is feasible, as the output corresponding to each user is merely a well defined set (usually a region of the output vector space) of scores -- for instance, the authentication mechanism decides that the input belongs to user 1 if the first component of the output exceeds a certain threshold. The attacker can then choose an output vector in this set, and work backwards towards the input of the ANN classifier. The recovered input of the classifier (note that usually a continuum of such valid inputs exist) can then be chosen as the biometric reading. Recall that under our threat model above, the attacker can force the authentication mechanism to consume the biometric reading of their choice.

In the case of the NFE-enhanced classifier, the chosen user's authentication record consists of a user name, a difference vector and a hash value. The difference vector contains no information about the pre-image of the hash value. Neither can such a pre-image be found from the hash value, as long as the hash function is pre-image resistant. Therefore, the attacker has no idea about what the output of the expander should be for the chosen user. The best the attacker can do is to choose randomly one of the codewords, and add to it the difference vector located in the chosen user's authentication record, in the hope that the result is actually the center of the chosen user's decision region. The attacker can then work back through the ANN+expander to produce a biometric reading corresponding to this result. Recall however that the codebook is designed such that the number of codewords is the same as the number of possible distinguishable user profiles. Therefore, choosing a codeword from the codebook at random is no better than choosing a person at random, and asking them to provide a biometric reading.  Now the probability that the attacker can successfully log in as a different user is equal to the probability that a non-legitimate random person can log in as that user by providing their own biometric readings, plus the probability that the attacker can find a pre-image for the user's hash value. Hence, within our threat model, and assuming a pre-image-resistant hash function, the attacker cannot win.

\paragraph*{NFEs codebooks, sphere packings, and the biometric entropy}

The purpose of the NFE is to cast the embeddings of different users' biometric readings in quasi-spherical regions in a certain vector space. In the left portion of Figure \ref{codebook} we show multiple such quasi-spherical clusters, such that each cluster (represented by markers with a specific shape and color) corresponds to the embeddings of the available biometric readings for a single user. In this figure, we decide to consider the support of the expander output as a sphere that includes all the available biometric data points. Taking the smallest such sphere may artificially reduce the perceived security of the biometric authentication system -- because it will result in a smaller codebook --  so, to be safe, consider a sphere of radius 10\% larger than that of the smallest outer sphere. 

The radius of the user-specific decoding region (the small sphere corresponding to the user of interest in Figure \ref{codebook}) is usually chosen to provide a certain false-positive-false-negative tradeoff. For simplicity of implementation, we choose the same radius (technically, the same error-correction code) for all users.

As an example, for the VGG-16 architecture retrofitted with the NFE, the FVC2006 database yields outer sphere of radius $1.0153$, and a small sphere of radius $0.7$ (this radius is chosen such that the distance-based decoder achieves 95.36\% accuracy on the training data set), meaning that at most $(1.0153/0.7)^{128}\simeq 4.7\cdot 10^{20}$ small spheres can be packed inside the large 128-dimensional sphere. We could say that the maximum number of distinct individuals whose biometric reading embeddings in our new vector space fit within non-overlapping small spheres is thus about $4.7\cdot 10^{20} \simeq 2^{68.67}$, or that the entropy of the secret biometric information, as represented in this space, is about $68.67$ bits. In other words, the proposed methodology for the expansion of biometric data provides -- as a byproduct -- a way of evaluating the authentication potential of various types of biometric data. It would be possible in this framework to decide (at least approximately, based on the upper and lower bounds on the entropy of a user's biometric data) whether fingerprint-based biometrics, for instance, are more or less secure than, say retina-scan-based biometrics. This evaluation is related to the average (across multiple possible AUs) accuracy of the associated classifier, but is not straightforward to derive from it, and would not be feasible in the absence of the expander.

We should note here that previous efforts to quantify the entropy of various types of biometrics take different approaches. For example, \cite{yankov2019fingerprint} produces empirical distributions of fingerprint data,  and yields a measure of entropy per pixel for different fingerprint databases. Their FVC2002 and FVC2004 databases are comparable to our FVC2006 database. However, their entropy upper bounds -- slightly above 0.25 bits per pixel -- would yield entropies in the order of 50,000 bits for each image. Note that this is the entropy of the fingerpint image, not the entropy of the fingerprint-based biometric modality that we calculate. To approximate the latter, \cite{yankov2019fingerprint} uses the mutual information between fingerprint images, and ends up with a maximum number of distinct individual representations of $10^{28}$ -- making our estimate of  $4.7\cdot 10^{20}$ somewhat conservative.
Other methods for evaluating biometric entropy are introduced in \cite{lim2015entropy} and \cite{sutcu2013biometric}.

%% file: classifier.tex
\section{An Instantiation  of NFE Retrofit}\label{demo}

To instantiate our NFE retrofit architecture described above, we had to make several design choices. First, we chose three pre-existing classifier architectures, as described in Section \ref{nn} below, as the basis for the retrofit. We had to slightly modify each one of the architectures. Next, we used an expander architecture consisting of multiple fully-connected layers, of decreasing sizes, to reduce the original classifier output size to a fixed size of 128 neurons. This was done to accommodate a tractable decoding mechanism. The error-correcting code used to provide the decoding mechanism was chosen to be a 128-dimensional low-density lattice code taken directly from \cite{sommer2008low}. The rest of this section contains additional details regarding our design choices.

\subsection{Classifier and Expander Architecture}\label{nn}

\paragraph*{VGG16} The VGG16 \cite{han2015deep,simonyan2014very} weights are pre-trained on ImageNet using Keras. To adapt this classifier to fingerprint-based user authentication, we removed the final softmax layer, rendering an output of size 4608. The size of the vector space (4608) is way too large to match to any efficient decoding mechanism. Therefore, to retrofit the NFE to this classifier, we constructed an expander, by adding three more fully connected layers of 512, 256 and 128 neurons respectively.

\paragraph*{ResNet50} The ResNet50 \cite{he2016deep} weights are pre-trained on ImageNet using Keras. After removing the final softmax layer, we were left with an output of size 2048 by average pooling the last layer. Then we constructed an expander by adding 4 fully connected layers of size 1024, 512, 256, and 128, respectively.

\paragraph*{MobileNet} We used a modified MobileNet v1 model architecture \cite{howard2017mobilenets} with weights pre-trained on ImageNet using Keras. After removing the final softmax layer, we were left with the output of 1024 by average pooling the last layer. Then we constructed an expander by adding 3 fully connected layers of size 512, 256 and 128, respectively.


\paragraph*{One Shot classification and Siamese network}
In the case of standard identification methods, a set of images are fed into an ANN to get an output probability for each one of the different classes. For example, if we want to distinguish between a cat and a dog we want to collect a lot of images (possibly more than 500 images per class) to improve model accuracy. The drawback of this type of network in fingerprint identification is \textbf{first}, it is nearly impossible to get a lot of images and \textbf{second}, if we want to include a new user in our database, we need to retrain the model to identify the new user as well. It is for these reasons that we choose to train our classifier in a Siamese network configuration, as explained below.  

It should be noted that for this specific application, our ``expander'' is in fact not expanding at all, but rather contracting the ANN's output. This is to reduce the complexity of the decoding involved in the secure sketch. Nevertheless, the same exact principles apply in situations in which the expander actually expands the output size.

A siamese network (sometimes called a twin neural network) is an ANN which learns to differentiate between two inputs instead of classifying. It takes two input images, runs through the same network simultaneously, and generates two vector embeddings of the images which are run through a logistic loss to calculate a similarity score between the two images \cite{bromley1994signature}. This is very useful as it does not require many data points to train the model. For training purposes, we only need to store one image belonging to the legitimate user as a reference image, and calculate the similarity for every new instance presented to the network.

\begin{figure}[]
	\centering
	\includegraphics[width=\linewidth]{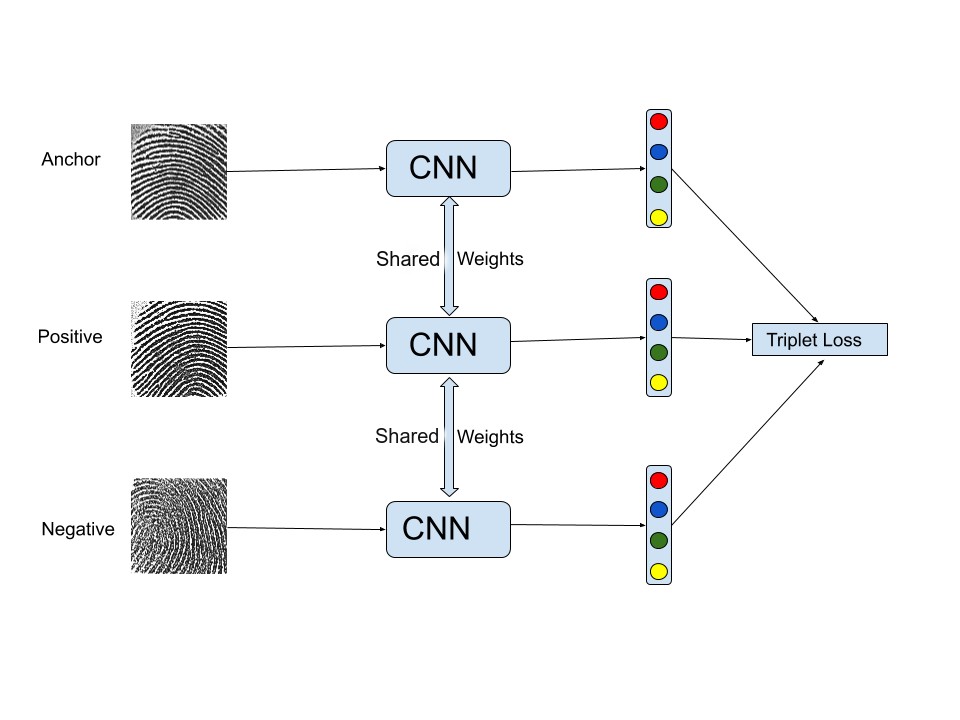}
	\caption{Triplet Loss architecture with Siamese network: Three images ("Anchor","Positive" and "Negative") are passed through the same CNN simultaneously to generate a final layer of 128 dimensional vector. Then all three vectors are passed through the triplet loss function to minimize the distance between "Anchor" and "Positive" as well as maximizing the distance between "Anchor" and "Negative".} \label{network}
	\vspace{-4mm}
\end{figure}

For our implementation, we used a triplet loss function with the Siamese networks. The benefit of using a triplet loss function (explained in the next subsection) in conjunction with a Siamese network is twofold \cite{dong2018triplet}: 
\begin{enumerate}
\item It extracts more features by learning to maximize the similarity between two similar images (Anchor-Positive) and the distance between two different images (Anchor-Negative) at the same time.
\item It generates more training samples than logistic loss. If we have $P$ similar pairs and $N$ dissimilar pairs then for logistic loss we will have $P+N$ total training samples. Whereas, we will have $PN$ triplets for training. This will impove the model accuracy.
\end{enumerate}
Our Siamese network architecture is depicted Figure \ref{network}. It is interesting to note that Siamese network training will naturally encourage the embeddings of the data points to cluster in spheres. This is a direct consequence of its loss function that causes embeddings from the same class to be close together, and embeddings from different classes to be well separated in terms of Euclidean distance. Nevertheless, under different circumstances, with more available training data, other expander architectures can be used, as long as their loss functions are adjusted to include similar distance-based penalties.

\paragraph*{Triplet Loss Function} Triplet loss functions are widely used in various applications in computer vision, such as face recognition \cite{DBLP:journals/corr/SchroffKP15}, person re-identification \cite{hermans2017defense} and image retrieval \cite{gordo2016deep}. Taking inspiration from that we used this for fingerprint verification. We will further explain how triplet loss functions work. 

\begin{figure}[]
	\centering
	\includegraphics[width=\linewidth]{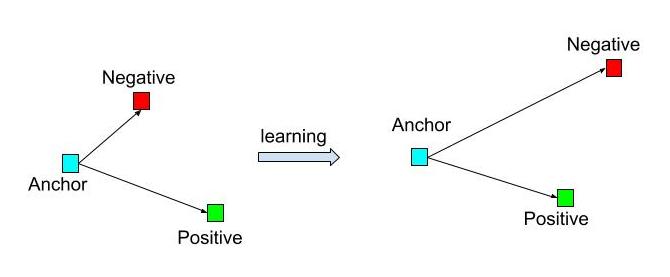}
	\caption{Triplet Loss architecture: The architecture tries to minimize the distance between "Anchor" and "Positive" and maximize distance between "Anchor" and "Negative"} \label{triplet loss}
	\vspace{-4mm}
\end{figure}

If we use a CNN to convert an image $x$ into a d-dimensional Euclidean space, then the embedding is represented by \(f(x) \in \mathbb{R}^d\). Here $f$ is the function computed by the CNN. For training we used triplets of fingerprint images, as shown in Figure \ref{triplet loss}:
\begin{itemize}
\item A is an "anchor" image -- a fingerprint image of a user.
\item P is a "positive" image -- a fingerprint image of the same user.
\item N is a "negative" image -- a fingerprint image of a different user.
\end{itemize}
We write triplets as $(A^{(i)},P^{(i)},N^{(i)})$ where i denotes the $i^{th}$ training example. We want to make sure that P is closer to A than N. Thus, we want
\[\|f(A^{(i)}) - f(P^{(i)})\|^2_2 + \alpha < \|f(A^{(i)}) - f(N^{(i)})\|^2_2\]
for all \(\{f(A^{(i)}),f(P^{(i)}),f(N^{(i)})\} \in T\), where $T$ is the set of all possible triplets.
Here, we want to make sure that the positive pair ($A^{(i)}-P^{(i)}$) has at least a margin difference of $\alpha$ over the negative pair ($A^{(i)}-N^{(i)}$). 
So the "triplet cost" function for the CNN becomes
\[\sum\limits_{i=1}^n\left[\|f(A^{(i)}) - f(P^{(i)})\|^2_2 - \|f(A^{(i)}) - f(N^{(i)})\|^2_2 + \alpha \right]_+.\]

Generating all possible triplets for training will result in slower convergence, 
so it is important to select a combination of "hard" and "easy" batches for the improvement of the model.

\subsection{The Error-Correction Code}
\begin{figure*}[]
	\centering
\begin{subfigure}[b]{0.49\textwidth}
         \centering
         \includegraphics[width=\textwidth]{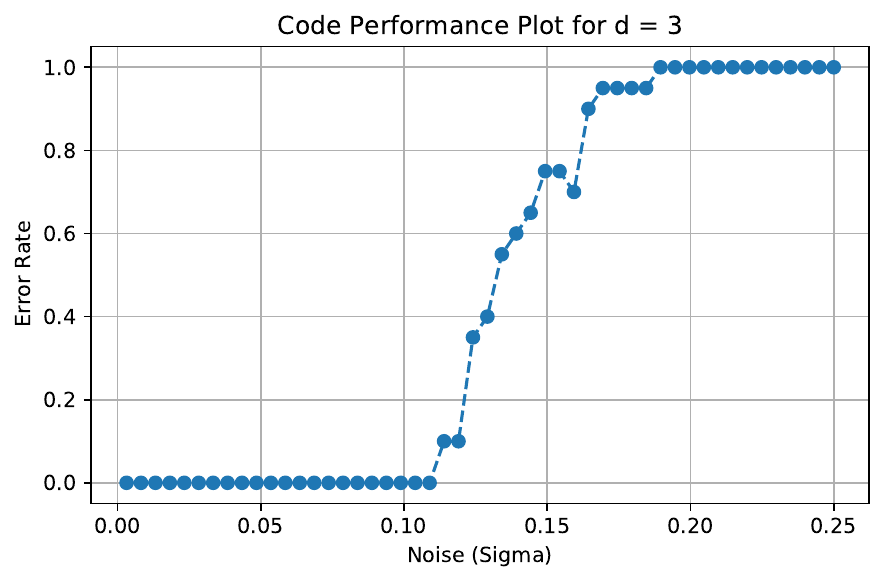}
         \label{coded3}
     \end{subfigure}
     \begin{subfigure}[b]{0.49\textwidth}
         \centering
         \includegraphics[width=\textwidth]{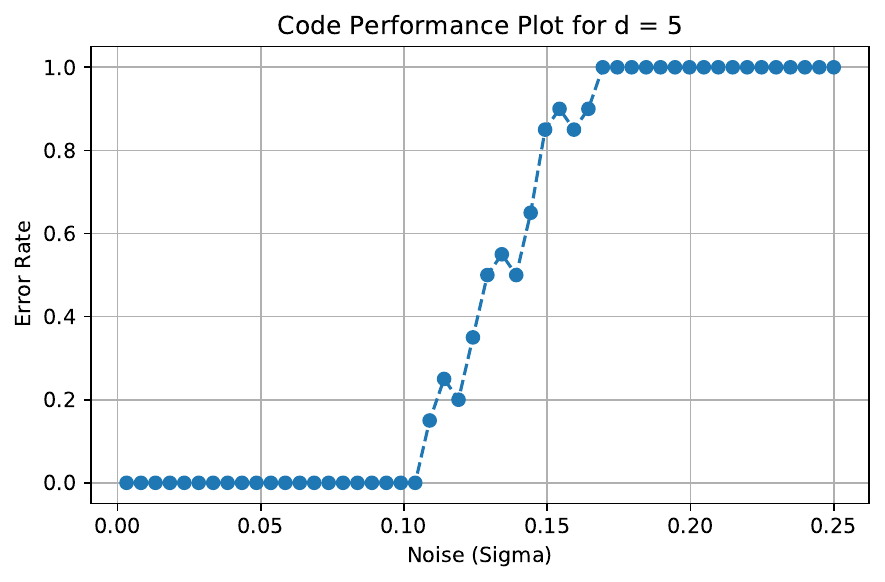}
         \label{coded5}
     \end{subfigure}
        \caption{Error rate vs. injected noise level for the Latin-Square LDLC with row/column degree $d$ of the parity check matrix.}
        \label{allCodeD3}
\end{figure*}

As mentioned in Section \ref{sketch}, for our error-correction code we used a low-density lattice code (LDLC), picked directly from \cite{sommer2008low}, in $n=128$ dimensions. As explained in \cite{sommer2008low}, the LDLC is completely defined by its sparse parity-check matrix $H$. To find an appropriate parity-check matrix, we used the \emph{Latin Square} construction of \cite{sommer2008low}, in which every row and every column of $H$ contains the same $d$ nonzero values. The $d$ nonzero values were selected as the first $d$ values in the sequence $\{1/2.31, 1/3.17,1/5.11,1.7.33,1/11.71\}$. The allocation of these values to the rows and columns of $H$ was done in accordance with the algorithms presented in Appendix VII of \cite{sommer2008low}. We experimented with $d=3$ and $d=5$. The error-correction performance of the two codes (for $d=3$ and $d=5$) are shown in Figure \ref{allCodeD3}, in which every point is obtained by applying the decoding algorithm 20 times  to a randomly-chosen codeword distorted by additive white Gaussian noise of standard deviation $\sigma$ (shown on the horizontal axis). We can see from the figures that the performance of the two codes is very similar in this artificial setting, with the $d=3$ code performing slightly better. This is also the case in practice, as illustrated next by the performance of the NFEs using the two codes.

We should note here the difference between binary codes and lattice codes. Binary error-correction codes use a codebook formed by a select few, sufficiently distant (usually in the sense of Hamming distance) binary sequences of a given length $n$, and all other binary sequences can be decoded to the closest sequence in the codebook. The rate of the binary code is controlled by the ratio $k/n$, where $k$ is the length of the message to be encoded (with $k<n$). The generator matrix of the code is thus a rectangular one, of size $n\times k$. One can therefore adjust the density of the codebook in the space $\{0,1\}^n$ by adjusting $k$ or $n$ or both. This helps when trying to fit a good codebook to a constellation of binary fingerprint representations from different users -- one can simply find the centroids of the users' fingerprint clouds, and produce a code with roughly the same density.

By contrast, LDLCs considered in this paper represent messages as vectors of $n$ integers, that are then mapped to codewords by a square $n\times n$ generator matrix, which is the inverse of the sparse (low-density) parity-check matrix. Decoding involves finding the vector of integers that is most likely to have generated the received sequence. Therefore, the information rate of LDLCs is only meaningful when an average power constraint is imposed on the codewords. This is not the case for our application. Instead, it was shown in \cite{poltyrev1994coding} that more meaningful performance metrics for infinite-constellation codes (like LDLCs), and ones that do not depend on power constraints, are directly related to the \emph{constellation density}. For our LDLC code, the constellation density is fixed. Thus, we cannot fit the codebook to the constellation of fingerprint representations -- instead, we have to fit the fingerprint constellation to the codebook. We do this by scaling all fingerprint representations by the same \emph{scale factor} $\gamma$. Choosing different scaling factors produces different false-positive rate (FPR) to false-negative rate (FNR) tradeoffs, thus enabling us to construct the receiver operating characteristic (ROC) curves.

Interestingly, a very similar scaling by $\gamma$ also enables us to produce different FPR-FNR tradeoffs when distance (norm) based decoding is used, by keeping the decoding distance threshold constant and varying $\gamma$.

\subsection{Some Guidelines for System Configuration}

Finally, before moving on to the performance and cost evaluation of our NFE implementation, we summarize several system configuration aspects of our NFE-retrofitted fingerprint-based biometric authentication, to serve as a quick reference for implementation.

\paragraph*{Number of users} There is no limit on the number of users in practice. Recall our (conservative, compared to \cite{yankov2019fingerprint}) estimate of a maximum of $4.7\cdot 10^{20}$ differentiable users. Clearly, if this many users were registered with the authentication server, then any input biometric would most likely correspond to at least one legitimate user. Fortunately, the world's population is still far below this number, and an authentication attempt would not only have to provide the biometric reading, but also the associated user name, making a successful authentication attempt very unlikely.

\paragraph*{Original classifier preparation} To prepare the original neural-network-based classifier for the NFE retrofit, we recommend removing the final softmax layer, if such a layer exists. The size of the output after this trimming procedure should be large enough to prevent the loss of biometric input information due to excessive compression.

\paragraph*{Codeword size} The appropriate codeword size $n$ depends on the efficiency of the decoding algorithm and on the acceptable computation overhead during user authentication. Based on the capabilities of consumer-grade hardware at the time of this writing, we recommend to choose $n$ in the range of $[100,500]$. See \cite{wang2019efficient} for more detailed results on the computational overhead as a function of $n$, for an efficient LDLC decoder. Also, as noted above, setting $d=3$ should provide very acceptable performance in this range.

\paragraph*{Setting the FPR-FNR tradeoff} As mentioned above, controlling the FPR-FNR tradeoff of the LDLC code is done not by modifying the code (whose constellation is fixed), but by scaling the NFE's output by a scalar factor $\gamma$. When implementing our NFE, we recommend trying multiple values of $\gamma$ to build an approximation of the ROC curve, and then choosing an operating point on this curve.

%% file: evaluation.tex
\section{NFE Evaluation}\label{eval}

To evaluate our NFE retrofit instance, we had to make several experimental design choices. First, we selected two databases of fingerprints, of different sizes. Next, we defined four different experiments, to allow us a comparison between the various retrofitted architectures, and two  different baseline architectures, for both databases, and all three classifiers. Finally, we experimented with multiple ways of tuning the pre-trained (and slightly modified, as explained in Section \ref{demo} above) classifiers to our databases, and chose the best-performing tuning modalities. The remainder of this section provides additional details on these design choices, as well as the experimental results.

\subsection{Datasets}

We used two datasets for our evaluation process: the 2006 Fingerprint Verification Competition (FVC2006) Database \cite{fvc2006} and the PolyU fingerprint database provided by Hong Kong Polytechnic University \cite{ployu2017}.

\paragraph*{The FVC2006 Database} This database consists of 4 distinct subsets DB1, DB2, DB3 and DB4. Each database consists of 150 fingers and 12 impressions per finger. Each subset is further divided into "set A" and "set B" where "set A" contains 140x12 images and "set B" contains 10x12 images  At the time of this writing we only used DB1, which has an image size of 96x96 but we expect similar results with the other databases. The images in the FVC2006 database were collected as part of the European Project BioSec, from anonymous volunteers, and was released publicly as part of the 2006 Fingerprint Verification Competition, organized by 
The Biometric System Laboratory (University of Bologna), the Pattern Recognition and Image Processing Laboratory (Michigan State University), the Biometric Test Center (San Jose State University) and the Biometrics Research Lab - ATVS (Universidad Autonoma de Madrid) \cite{fvc2006}. At the time of this writing, we have no reason to doubt the ethical aspects of the data collection process.

\paragraph*{The PolyU Database} This database contains 2016 fingerprint images from 336 different users, i.e., 6 impressions per user. The size of the image is 328x356 initially. To make it consistent with our initial dataset, we augmented the images using several data augmentation techniques like the addition of random noise and random rotation. In this way, we generated 6 additional impressions per user, bringing our dataset to 12 fingerprint images per 336 users and converted the size of these images to 224x224. The PolyU database was released in 2017 for public use by the The Hong Kong Polytechnic University. At the time of this writing, we have no reason to doubt the ethical aspects of the data collection process.

\paragraph*{Training Data} For either one of the databases, training data consisted of 10 impressions per finger (total of 140x10 images for the FVC2006 database, and a total of 336x10 images for the PolyU database). Out of these images, we generated the triplet pairs. In this paper, we used 50\% "hard" and 50\% "easy" triplet pairs.

\paragraph*{Testing Data} For either of the two databases, testing data consisted of 2 impressions per finger.


\subsection{Experimental Setup}
We conducted four experiments using ResNet50, MobileNet and VGG16 architectures with two different datasets FVC2006 and PolyU.

\paragraph*{Baseline Classifier (BC)}
Our first experiment is meant to provide a baseline for the evaluation of the expander-enhanced architectures. For this purpose, we trained the three classifier architectures (ResNet50, MobileNet and VGG16) separately. Since we used two datasets -- FVC2006 with 140 users and PolyU with 336 users -- our classification model’s top layer was a softmax layer with 140 nodes in the case of FVC2006 and 336 nodes in the case of the PolyU dataset, respectively.

The final layer of the classification model gives us the probabilities (or, more generally, some scores for the events) that the input image belongs to each possible user. We transform the model into a binary classifier by establishing a score threshold for each one of the users. Hence, to calculate the false positive rate for a single class, we took 20 different fingerprint images from other classes fed them through the network. If the score output value for the image is greater than the probability threshold for that particular class we are interested in, then that’s the condition of false positive.

Similarly, to calculate the FNR for a single class, we took 2 images for that class from the test set, augmented them using several data augmentation techniques like addition of random noise and rotation to generate another 20 images and fed them through the network. If the output probability value of the node of the class that we are concerned with is less than the probability threshold, then that’s the condition of false negative.

\paragraph*{Classifier Trained in a Siamese-Network Configuration (CSN):}
In the second experiment, we trained the three classifiers individually,  using a Triplet-based Siamese-network configuration. No expander was used in this experiment, and thus the size of the final embeddings vary according to the architecture: ResNet50 uses an embedding of size 2048 (the size of the final layer before the softmax layer), MobileNet uses an embedding size of 1024, and VGG-16 uses a size of 4608. This experiment was designed to provide a fair baseline comparison for the expander-enhanced architectures, just in case that this Siamese-network configuration training proved overall superior to the standard training of experiment BC above -- it turns out that its superiority is in fact classifier-dependent.

\paragraph*{Classifier Trained in a Siamese-Network Configuration Plus Expander (CSN+ESN)}
In the third experiment, we use the classifiers already-trained under the second (CSN) experiment, and retrofit them with expanders. Then, only the expanders are trained, using the classifier-plus-expander architectures in the standard Siamese-network configuration. This experiment emulates a scenario in which an existing neural network-based classification architecture is already trained and provides good performance by itself. In such a situation, training only the expander while keeping the original classifier's weights fixed can provide significant computational savings.

\paragraph*{Jointly Trained Classifier and Expander ((C+E)SN)}
In this last experiment, we first retrofitted all three classification architectures with an expander, and trained the classifier-plus-expander architectures from scratch, using the standard Siamese-network configuration. Our initial expectation was that such joint training of the classifier and expander would provide the best performance. In reality, the results show that this is in fact classifier-dependent.

For each of the CSN+ESN and (C+E)SN experiments, we execute, and report the results of, two types of error-correction decoding: (a) using simple fixed-distance-based decoding, and (b) using the LDLC decoder. The purpose of the distance-based decoder is solely to provide a baseline for evaluating the performance degradation due to the LDLC decoder (however, in reality, as our results will show, the LDLC decoder usually out-performs the distance-based decoder). The distance-based decoder is not a viable option in practice, as it would require that the center of the AU's decision region is saved as part of the AU's authentication record. This would defeat the purpose of the NFE.

\subsection{Training of the Classifiers}

\paragraph*{VGG-16} For VGG-16 architecture, we trained all its layers for the both baseline and the Siamese-network-based configurations, to get the better generalization of the model.
\paragraph*{ResNet50}
For ResNet50 architecture, in the Siamese-network-based configuration, we froze the first 143 layers and trained the remaining 32 layers to perform the transfer learning. However, in the case of the baseline model, we trained the whole architecture without freezing any layers.
\paragraph*{MobileNet}
For the MobileNet architecture in the Siamese-network-based configuration, we trained just the last 23 layers. However, in the case of the baseline model, we trained all the layers of the architecture to get a better generalization of the model.

\subsection{Results}
We expect that our NFE retrofit contains two sources of performance degradation, and we proceed to systematically investigate each. The first potential source of performance degradation is the injection of the expander into the system's architecture. This should be especially critical when the expander in fact reduces the output size of the original architecture, rather than expanding it. This is exactly the case with our implementation, where the output size is reduced from 4608, 2048 or 1024 to 128. 

The second source of performance degradation is the use of an error-correction code in place of distance-based decoding. This step is essential to the security of the NFE architecture, as it enables automatic decoding in the absence of a class representative (or centroid). Such a representative is required by distance-based decoding, and constitutes a severe security vulnerability. 

\paragraph*{The effect of the expander}
To evaluate the impact of the expander, we compared the performance of the three classifier architectures in the BC and CSN experiments, to the NFE-retrofitted architectures -- experiments CSN+ESN and (C+E)SN -- but performing distance-based decoding instead of LDLC-based decoding. For our two datasets, the results are given in Tables \ref{table:fvc2006EER} and \ref{table:polyuEER}.

\begin{table}[!htbp]
\begin{tabular}{p{1.3cm}|p{0.55cm}|p{0.85cm}|p{0.55cm}|p{0.85cm}|p{0.55cm}|p{0.85cm}}\hline
    \multirow{2}{*}{Experiment} & \multicolumn{2}{c|}{ResNet} & \multicolumn{2}{c|}{MobileNet} & \multicolumn{2}{c}{VGG-16}\\ \cline{2-7}
        &{\scriptsize EER} & {\scriptsize AUROC} & {\scriptsize EER} & {\scriptsize AUROC} & {\scriptsize EER} & {\scriptsize AUROC}\\ 
    \hline
    BC & 0.032 & 0.972 & 0.022 & 0.969 & 0.040 & 0.940 \\
    \hline
    CSN & 0.065 & 0.982 & 0.055 & 0.989 & 0.014 & 0.999 \\
    \hline
    CSN+ESN distance & 0.054 & 0.986 & 0.044 & 0.994 & 0.013 & 0.990 \\
    \hline
    CSN+ESN LDLC & 0.053 & 0.987 & 0.025 & 0.995 & 0.013 & 0.999 \\
    \hline
    (C+E)SN distance & 0.066 & 0.984 & 0.042 & 0.993 & 0.020 & 0.997 \\ 
    \hline
    (C+E)SN LDLC & 0.065 & 0.984 & 0.039 & 0.993 & 0.020 & 0.995 \\ \hline
    \end{tabular}
    \caption{Equal error rates (EER) and areas under the receiver operating characteristics (ROC) curve (AUROC) for our four experiments, under each of the three architectures, when using the FVC2006 database.}
    \label{table:fvc2006EER}
\end{table}

We notice that for the FVC2006 database (Table \ref{table:fvc2006EER}), with small exceptions, all three architectures exhibit comparable EER values under the \emph{CSN} experiment, with the \emph{CSN+ESN distance} and \emph{(C+E)SN distance} experiments. However, it appears that the ResNet50 and MobileNet architectures in the \emph{BC} experiment exhibit significantly lower EER values, albeit also smaller AUROC values, than in the other two experiments. This shows that some particular performance degradation around the EER point may be caused by the training configuration, i.e. training under the baseline configuration outperforms training under the Siamese-network-based configuration. However, considering the AUROC values, such performance degradation is localized around the EER point.

The opposite holds true for the VGG-16 architecture, where the BC experiment shows higher EER  -- and this time also lower AUROC -- than all the other experiments.
\begin{table}[!htbp]
\begin{tabular}{p{1.3cm}|p{0.55cm}|p{0.85cm}|p{0.55cm}|p{0.85cm}|p{0.55cm}|p{0.85cm}}\hline
    \multirow{2}{*}{Experiment} & \multicolumn{2}{c|}{ResNet} & \multicolumn{2}{c|}{MobileNet} & \multicolumn{2}{c}{VGG-16}\\ \cline{2-7}
        &{\scriptsize EER} & {\scriptsize AUROC} & {\scriptsize EER} & {\scriptsize AUROC} & {\scriptsize EER} & {\scriptsize AUROC}\\ 
    \hline
    BC & 0.031 & 0.958 & 0.017 & 0.979 & 0.062 & 0.972 \\
    \hline
    CSN & 0.035 & 0.995 & 0.025 & 0.996 & 0.013 & 0.999 \\
    \hline
    CSN+ESN distance & 0.037 & 0.992 & 0.026 & 0.995 & 0.007 & 0.999 \\
    \hline
    CSN+ESN LDLC & 0.037 & 0.992 & 0.025 & 0.995 & 0.007 & 0.999 \\
    \hline
    (C+E)SN distance & 0.059 & 0.985 & 0.054 & 0.986 & 0.033 & 0.993 \\ 
    \hline
    (C+E)SN LDLC & 0.058 & 0.985 & 0.054 & 0.981 & 0.034 & 0.981 \\ \hline
    \end{tabular}
    \caption{Equal error rates (EER) and areas under the receiver operating characteristics (ROC) curve (AUROC) for our four experiments, under each of the three architectures, when using the PolyU database.}
    \label{table:polyuEER}
\end{table}

The same difference in performance between the \emph{BC} and \emph{CSN} experiments is observed for the PolyU dataset (Table \ref{table:polyuEER}), where again it appears that ResNet50 and MobileNet do significantly better in the BC experiment, while VGG-16 does significantly worse.

Interestingly, on both datasets, all architectures do much better in the \emph{CSN+ESN distance} experiment than in the \emph{(C+E)SN distance} experiment, suggesting that training of the expander alone, rather than joint training of the expander and original classifier, is the better choice.

\paragraph*{The effect of the decoder}
To evaluate the impact of the error-correction-code on the NFE performance, we compare the results in the experiments \emph{CSN+ESN distance} and \emph{(C+E)SN distance} to the experiments \emph{CSN+ESN LDLC} and \emph{(C+E)SN LDLC}, respectively. We notice that for both datasets, and for each one of the classifiers, the EER and AUROC values appear very similar between the distance-based decoding and the LDLC-based decoding. If anything, the LDLC-based decoder seems to perform very slighly better. One notable exception is the case of the MobileNet architecture, on the FVC2006 dataset (Table \ref{table:fvc2006EER}), where the LDLC decoder appears to help a lot, lowering ERR from 4.4\% (for the case of \emph{CSN+ESN} distance-based decoding) to 2.5\%. Such improvements in performance due to LDLC decoding may be explained by the fact that the LDLC decoding regions are not perfectly spherical (even in 128 dimensions), and fill up the vector space better than perfectly spherical regions (which are implicit with distance-based decoding).

\paragraph*{Overall effect of the NFE retrofit}
Connsidering the best performance among the \emph{BC} and \emph{CSN} experiments, and the best performance among the \emph{CSN+ESN LDLC} and \emph{(C+E)SN LDLC} experiments, we draw the conclusion that there appears to be no, or very slight, performance degradation due to the use of the NFE. This is an encouraging result, and should motivate the future evaluation of the application of NFEs to other architectures, and to other biometric authentication modalities. 

We can also notice that overall, the PolyU Dataset performed better than the FVC2006 dataset in most scenarios. This might be because PolyU images are of bigger resolution and have less noise than FVC2006 images.

\paragraph*{Training and running times}

We evaluated the training times and the running times of the three different architectures, under each one of our four experiments and two datasets. All our models were trained on a GeForce RTX 2080 TI GPU with 4 cores, with the maximum memory size at 25GB. For studying the runtime performance, we ran our user authentication mechanisms on a more realistic work station -- specifically, a personal laptop with Apple’s M1 silicon chip, and with 16GB of memory.

The execution times are listed in Tables \ref{table:fvc2006time} and \ref{table:polyutime} for the FVC2006 and the PolyU datasets, respectively. For training, we list in parentheses the number of training epochs that were necessary for each one of the experiments to observe the convergence of the training process. We note that for the BC experiment, convergence is observed a lot sooner than for the other experiments (fewer than 80 epochs, compared to over 4000 epochs). This is most probably due to the Siamese-Network configuration, which is employed by all the other experiments.

\begin{table}[!htbp]
\begin{tabular}{p{1.3cm}|p{1cm}|p{1.25cm}|p{1.25cm}|p{1.25cm}}\hline
    \multicolumn{2}{c|}{Experiment} & ResNet50 & MobileNet & VGG-16\\ 
    \hline
    \multirow{2}{*}{BC} & training & 7m11s {\scriptsize (60 ep)} & 7m14s {\scriptsize (80 ep)} & 8m48s {\scriptsize (70 ep)} \\ \cline{2-5}
                        & runtime  & 70ms   & 48ms  & 55ms \\
    \hline
    \multirow{2}{*}{CSN} & training & 1h53m22s {\scriptsize (5000 ep)} & 1h06m12s {\scriptsize (5000 ep)} & 1h05m39s {\scriptsize (5000 ep)}\\ \cline{2-5}
                        & runtime  & 529ms   & 264ms  & 173ms \\
    \hline                   
    \multirow{2}{*}{CSN+ESN} & training & 1h50m20s {\scriptsize (5000 ep)} & 1h08m13s {\scriptsize (5000 ep)} & 1h00m41s {\scriptsize (5000 ep)} \\ \cline{2-5}
                        & runtime  & 645ms   & 343ms  & 134ms \\
    \hline             
    \multirow{2}{*}{(C+E)SN}& training & 1h57m00s {\scriptsize (5000 ep)} & 1h06m30s {\scriptsize (5000 ep)} & 1h08m48s {\scriptsize (5000 ep)} \\ \cline{2-5}
                        & runtime  & 589ms   & 325ms  & 189ms \\  \hline
    \end{tabular}
    \caption{Training and running times for our four experiments, under each of the three architectures, when using the FVC2006 database.}
    \label{table:fvc2006time}
\end{table}

We observe that under the BC experiment, both training and running times are significantly smaller than in all the other experiments. We also note that there are only small differences between the \emph{CSN+ESN} and \emph{(C+E)SN} experiments, and the \emph{CSN} experiment, in both training and running times, and with all three classifier architectures. As expected, training and running for the \emph{CSN+ESN} and \emph{(C+E)SN} experiments in general takes slightly longer, but not by much.

We can therefore infer that the addition of the expander incurs minimal overhead. However, switching from a standard baseline architecture to a Siamese-network-based architecture does incur significant penalties -- in the order of a few hours for training, and in the order of hundreds of milliseconds for runtime.
\begin{table}[!htbp]
\begin{tabular}{p{1.3cm}|p{1cm}|p{1.25cm}|p{1.25cm}|p{1.25cm}}\hline
    \multicolumn{2}{c|}{Experiment} & ResNet50 & MobileNet & VGG-16\\ 
    \hline
    \multirow{2}{*}{BC} & training & 27m50s {\scriptsize (50 ep)} & 21m38s {\scriptsize (40 ep)} & 22m35s {\scriptsize (50 ep)} \\ \cline{2-5}
                        & runtime  & 81ms   & 55ms  & 63ms \\
    \hline
    \multirow{2}{*}{CSN} & training & 3h12m21s {\scriptsize (4000 ep)} & 2h10m08s {\scriptsize (4000 ep)} & 3h26m43s {\scriptsize (4000 ep)}\\ \cline{2-5}
                        & runtime  & 579ms   & 214ms  & 151ms \\
    \hline                   
    \multirow{2}{*}{CSN+ESN} & training & 3h17m23s {\scriptsize (4000 ep)} & 2h08m06s {\scriptsize (4000 ep)} & 3h09m32s {\scriptsize (4000 ep)} \\ \cline{2-5}
                        & runtime  & 459ms   & 227ms  & 160ms \\
    \hline             
    \multirow{2}{*}{(C+E)SN}& training & 3h12m45s {\scriptsize (4000 ep)} & 2h08m46s {\scriptsize (4000 ep)} & 3h22m00s {\scriptsize (4000 ep)} \\ \cline{2-5}
                        & runtime  & 755ms   & 240ms  & 156ms \\  \hline
    \end{tabular}
    \caption{Training and running times for our four experiments, under each of the three architectures, when using the PolyU database.}
    \label{table:polyutime}
\end{table}

We should note here that the running times reported in Tables \ref{table:fvc2006time} and \ref{table:polyutime} do not include the running times of the decoder for the \emph{CSN+ESN} and \emph{(C+E)SN} experiments. Our implementation of the LDLC decoder is based directly on the method proposed in \cite{sommer2008low}, which is known to be rather inefficient. Many subsequent works have proposed much more efficient techniques for decoding LDLCs, see for example \cite{yona2009efficient,bickson2009low,liu2019efficient,wang2019efficient}. For example, when considering decoding parameters similar to ours ($d=3$ and $n\in [100,1000]$), running on consumer-grade hardware, \cite{wang2019efficient} reports running times of around $4.2ms$ per iteration for $n=100$ and $36 ms$ per iteration for $n=1000$, leading to total decoding times of between $0.42s$ and $3.6s$ (corresponding to 100 iterations). These decoding times are well within the acceptable range for user authentication. Nevertheless, the integration of efficient LDLC decoders with NFEs is outside the scope of the current paper, and is the subject of future work.

\paragraph*{Comparison with plain fuzzy extractors}

A performance comparison with the plain fuzzy extractors of \cite{dodis2004fuzzy} is not fair -- it has already been established that they suffer from low performance \cite{maio2002fvc2000,arakala2007fuzzy}. However, in terms of security and cost comparisons, plain fuzzy extractors appear similar to our proposed NFEs. The security provided by NFEs relies on the secure sketch construction -- a concept borrowed from the plain fuzzy extractors, and hence the security guarantees are identical. Since both types of extractors have to store only hash values and difference vectors for each of the registered users, the storage requirements are also similar -- of course, NFEs have to also store the weights of the associated neural network. In terms of runtime performance, both NFEs and plain fuzzy extractors have to implement a complex decoding procedure, which accounts for the bulk of the user authentication computation in NFEs, and for almost the entire computation in the case of plain fuzzy extractors.

%% file: conclusion.tex
\section{Conclusions}\label{conclusions}

In this paper, we presented a secure architecture for biometric user authentication (or identification), which avoids the storage of information (such as neural network weights or specific biometric data) that could be used by malicious entities for constructing artificial biometric inputs able to pass the authentication tests. To that extent, the proposed architecture aligns with the current paradigm for handling users' passwords. The proposed architecture -- which we call a \emph{neural fuzzy extractor} -- works by coupling the classifier with a fuzzy extractor -- this is possible by the use of an \emph{expander}, which is a neural network that can be trained at the same time as, or after, the classifier. The NFE architecture can be retrofitted to any already-existing well-performing classifier, and should combine the classification performance of artificial neural networks with the security of fuzzy extractors. Our instantiation of the NFE for fingerprint-based authentication demonstrates how a pre-existing classifier can be retrofitted for NFE implementation, and how two different types of training can be conducted for NFE-specific output-space sphere clustering. Our experimental results show that the addition of the NFE does not incur significant performance degradation (in terms of equal-error rates and areas under the receiver-operating-characteristics curves), but does incur higher training and runtime overhead, mainly due to the Siamese-network-based configuration. Nevertheless, we believe that the 1 to 3 hour increases in training times is well worth the security advantage provided by NFEs, while the 100 to 700 millisecond increases in running times is well within the acceptable range for applications.  Future work will focus on (1) studies of the application of NFEs to other types of biometrics and (2) the construction and training of expanders suited to coding over binary extension fields.  

%% file: acknowledgement.tex
\section{Acknowledgements}

This material is based upon work supported by the National Science Foundation under Grant No.  1619201.